\documentclass[preprint,showpacs,preprintnumbers,amsmath,amssymb,aps]{revtex4-1}
\usepackage{graphicx,dcolumn,bm}
\usepackage{amssymb,amstext,amsmath}

\usepackage{graphicx}
\usepackage{dcolumn}
\usepackage{bm}
\usepackage{amssymb}
\usepackage{multirow}
\usepackage{bigstrut}
\usepackage{makecell,rotating}

\usepackage{mathrsfs}
\usepackage{booktabs}
\usepackage{threeparttable}
\usepackage{multirow}
\usepackage{subfigure}
\usepackage{epsfig}
\usepackage{threeparttable}
\usepackage{chngpage}
\usepackage{float}

\usepackage{amstext}
\usepackage{amsmath}

\begin{document}

\title{Aspiration dynamics generate robust predictions in structured populations}

\author{Lei Zhou$^{1}$, Bin Wu$^{2}$, Jinming Du$^{3,4}$, and Long Wang$^{1, }$}
\email{Corresponding author.}
\affiliation{
$^{1}$Center for Systems and Control, College of Engineering, Peking University, Beijing 100871, China \\
$^{2}$School of Sciences, Beijing University of Posts and Telecommunications, Beijing 100876, China \\
$^{3}$Liaoning Engineering Laboratory of Operations Analytics and Optimization for Smart Industry, Northeastern University, Shenyang 110819, China \\
$^{4}$Institute of Industrial and Systems Engineering, Northeastern University, Shenyang 110819, China \\
}

\begin{abstract}
Evolutionary game dynamics in structured populations are strongly affected by updating rules.
Previous studies usually focus on imitation-based rules, which rely on payoff information of social peers.
Recent behavioral experiments suggest that whether individuals use such social information for strategy updating may be crucial to the outcomes of social interactions.
This hints at the importance of considering updating rules without dependence on social peers' payoff information, which, however, is rarely investigated.
Here, we study aspiration-based self-evaluation rules, with which individuals self-assess the performance of strategies by comparing own payoffs with an imaginary value they aspire, called the aspiration level.
We explore the fate of strategies on population structures represented by graphs or networks.
Under weak selection, we analytically derive the condition for strategy dominance, which is found to coincide with the classical condition of risk-dominance.
This condition holds for all networks and all distributions of aspiration levels, and for individualized ways of self-evaluation.
Our condition can be intuitively interpreted: one strategy prevails over the other if the strategy brings more satisfaction to individuals than the other does.
Our work thus sheds light on the intrinsic difference between evolutionary dynamics induced by aspiration-based and imitation-based rules.
\end{abstract}
\maketitle

\section{Introduction}
Past decades have seen intensive investigations of evolutionary games in structured populations \cite{Nowak1992, Hauert2004, Szabo2007, Nowak2010, Debarre2014}.
One of the most important questions is how population structure alters evolutionary outcomes.
It is shown that this strongly depends on updating rules \cite{Szabo2007, Roca2009Temporal, Tarnita2009, Allen2014}.
Update rules are explicit behavioral rules of individuals, which specify what kind of information they use and how they process such information to determine future behaviors \cite{Szabo2007}.
In evolutionary games, the information required by updating rules usually includes individuals' strategies and payoffs.
As the input for decision making, the information used is likely to affect individuals' behavioral updating, resulting in changes at the population level.
Indeed, recent human behavioral experiments suggest that whether individuals use social peers' payoff information to update behavior may be crucial to the outcome of social interactions, for example, the level of cooperation in groups \cite{Molleman2014, Van2015} or on network-structured populations \cite{Grujic2010, Gracia2012, Grujic2014, Rand2014}.

Based on the relevance of social peers' payoffs, updating rules in theoretical models can be classified into two classes: imitation-based (require) and self-evaluation based (not require).
Under imitation-based rules, individuals update strategies by copying strategies of the more successful peers.
When using self-evaluation based rules, individuals self-assess performance of strategies and then switch to strategy alternatives \cite{Szabo2007}.
Self-evaluation can be based on aspiration: individuals compare achieved payoffs with their endogenous aspirations and then switch based on the shortfall of payoffs \cite{Karandikar1998, Posch1999, Bendor2011}.
Updating rules of these two classes are both common in practice and they are tailored for different environment.
For example, if individuals are not confident to make decisions or uncertain about the consequences, imitating the more successful provides valuable shortcuts for decision-makers.
Self-evaluation, instead, is efficient and superior when social information is unavailable, regarded as unreliable, or costs individuals too much to gather and process.
Albeit their prevalence, most studies in structured populations focus on imitation-based rules \cite{Nowak1992, Hauert2004, Santos2006, Traulsen2006b, Ohtsuki2006, Szabo2007, Tarnita2009, Roca2009Temporal, Nowak2010, Allen2014, Debarre2014, Allen2017}, whereas self-evaluation based rules receive much less attention \cite{Chen2008, Roca2009BR, Roca2011, Du2014, Du2015}.
Since evolutionary outcomes are highly sensitive to updating rules, a lack of sufficient investigation in self-evaluation based rules may prevent us from better understanding how population structure affects the fate of strategies.

Indeed, theoretical studies find that self-evaluation based rules lead to intrinsically different evolutionary outcomes from imitation-based ones \cite{Chen2008, Roca2009BR, Du2014, Du2015}.
One of such differences is the effect of population structure on the fate of strategies: under aspiration-based self-evaluation rules, the fate of strategies on regular and random graphs is found to be the same as that in well-mixed populations \cite{Du2015}, whereas imitation-based rules lead to outcomes sensitive to population structures and other model details \cite{Tarnita2009, Allen2014}.

As yet, it remains unclear how evolutionary games under aspiration-based self-evaluation rules (for short, aspiration dynamics) affects the fate of strategies on general population structures.
Here, we consider aspiration dynamics on any graphs or networks.
Under the limit of weak selection, we analytically derive a condition for one strategy to prevail over the other, which is found to coincide with the classical condition of risk-dominance.
This condition holds for every network and every distribution of aspiration levels, and for individualized ways of self-evaluation.
If aspirations are differentiated by strategies, we find that the condition of risk-dominance is altered and cooperation can evolve in the Prisoner's Dilemma game if defectors aspire more than cooperators.
The intuitive interpretation of our results is as follows: one strategy prevails over the other if the strategy brings more satisfaction to individuals than the other does.
Our work reveals that switching off from the payoff information of social peers when updating strategies has a nontrivial impact on the evolutionary outcomes.

\section{Model}
We consider a population with fixed size $N$ ($N\ge 2$).
The population structure is depicted by a static weighted graph with edge weights $w_{ij}$, where vertices represent individuals and edges indicate who interacts with whom.
Self-interactions are excluded.
Individuals collect edge-weighted average payoffs by playing games with their nearest neighbors \cite{Allen2017}.
The total number of interactions each individual $i$ engages in is $d_i=\sum_{j=1}^N w_{ij}$ ($i=1,2,\cdots,N$).
We require $d_i>0$ for all $i$, which means that each individual at least has one neighbor to interact with.
Visually, the graph should have neither isolated vertices nor self-loops, which are natural assumptions when studying evolutionary games on graphs.
In each game, individuals can play either strategy $A$ or strategy $B$.
The payoff matrix of the game is given by
\begin{align}\label{payoffMatrixOriginal}
\bordermatrix{
  & A & B \cr
A & a & b \cr
B & c & d \cr},
\end{align}
where both players get payoff $a$ if they play strategy $A$ ($A$-player) and get $d$ if they play strategy $B$ ($B$-player); if an $A$-player encounters a $B$-player, the former obtains payoff $d$ and the latter $c$.
For each individual $i$, we denote $\pi_{i,X}$ as its payoff when it uses strategy $X$ ($X=A,B$).

At each time step, an individual is randomly selected and given the opportunity to revise its strategy.
We assume individuals follow self-evaluation based rules, under which they evaluate their strategies by comparing the realized payoffs with their endogenous aspirations.
Aspirations are either personalized \cite{Wu2018}, which means each individual $l$ has its own aspiration $\alpha_l$ ($l=1,2,\cdots, N$), or contingent on strategies, which means that individuals using strategy $A$ have an aspiration $\alpha_A$ and those using $B$ have $\alpha_B$.
For the sake of simplicity, we consider fixed aspirations, which means there is no adaptation of aspirations due to learning.
If such aspiration-driven updating rules is deterministic, the aspiration level serves as a sharp boundary between satisfaction and disappointment \cite{Simon1959}: if an individual's payoff exceeds its aspiration, the outcome is deemed as satisfactory and it will repeat its strategy; if the payoff is otherwise lower than the aspiration, it feels disappointed and will switch to the other strategy.
In real-life situations, strategy updating involves mistakes and admits bounded rationality, which allows individuals to switch strategies probabilistically (stochastically).
The probability can be determined by the level of satisfaction, i.e., the difference between the achieved payoff and aspiration level.
In our model, we use updating functions \cite{Wu2013, Wu2018} to map the payoff-aspiration difference into the switching probability.
We allow individuals to have their own updating functions since they may behave differently even for the same level of satisfaction.
Albeit this flexibility, all updating functions should ensure that individuals have decreasing tendency to switch a strategy if it brings more satisfaction.

\begin{figure}
  \centering
  \includegraphics[width=0.8\textwidth]{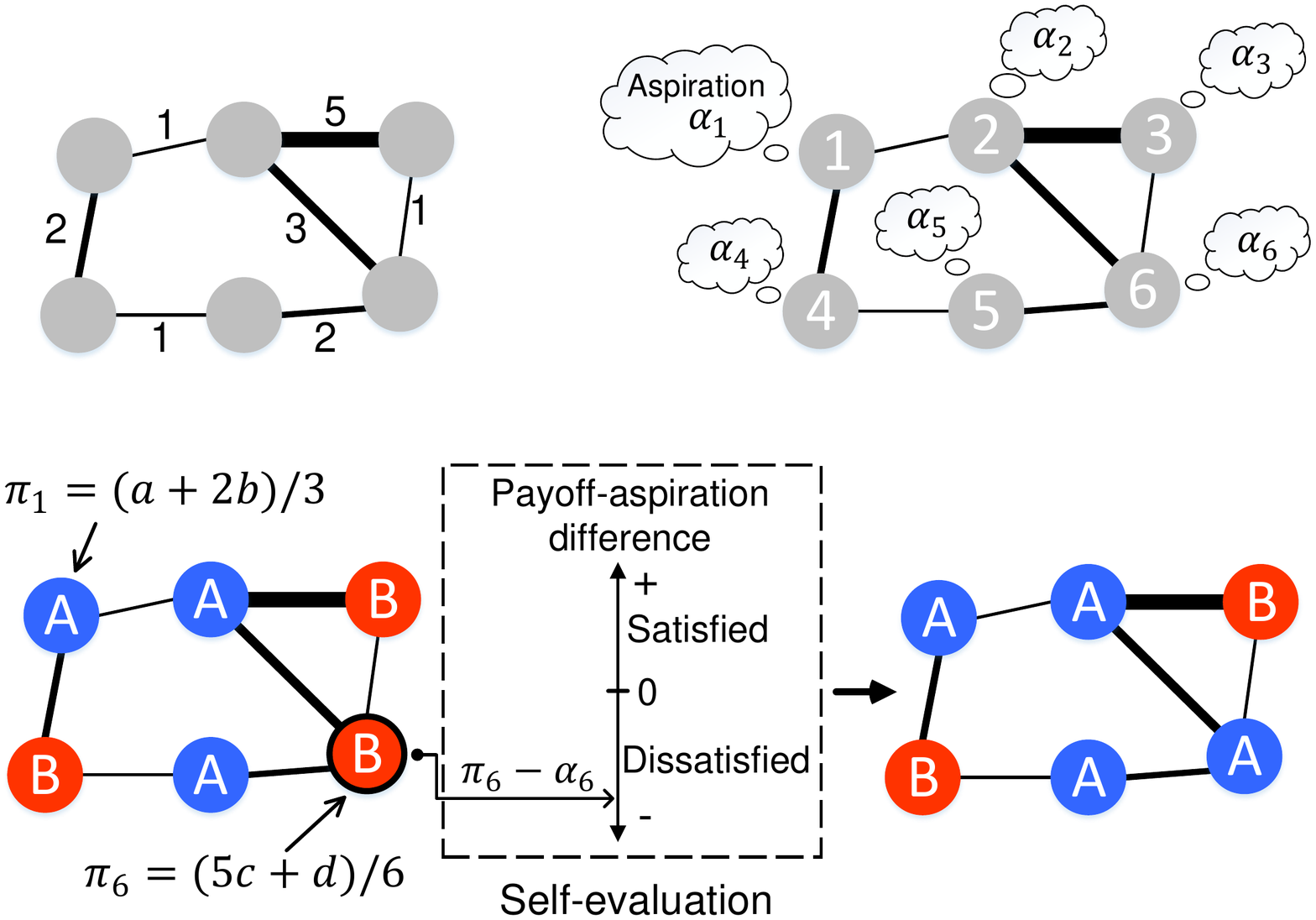}\\
  \caption{Aspiration dynamics on weighted graphs.
  Upper left panel: An undirected weighted graph with edge weight $w_{ij}$.
  Upper right panel: Individuals occupy vertices of the graph and each individual $l$ has an imaginary payoff value $\alpha_l$ they aspire, called aspiration level.
  Lower panel: For aspiration dynamics, at each time step, an individual is randomly selected (marked by black circle).
  It garners edge-weighted average payoffs ($\pi_6$) by playing games with its nearest neighbors \cite{Allen2017}.
  Then it self-evaluates the performance of the strategy in use by calculating the payoff-aspiration difference ($\pi_6-\alpha_6$).
  If the payoff exceeds the aspiration, it feels satisfied and is more likely to keep its current strategy; otherwise, it is prone to switch.
  As illustrated, $\pi_6-\alpha_6<0$ and the corresponding individual switches from strategy $B$ to $A$.}\label{fig1}
\end{figure}

Here, we employ stochastic self-evaluation based rules.
Under stochastic rules, the strictness of the strategy evaluation, namely, how much the payoff-aspiration difference affects individuals' decision-making, is controlled by the selection intensity $\beta \ge 0$ \cite{Nowak2004, Traulsen2007}.
The resulting aspiration dynamics admit a stationary distribution, which is similar to the mutation-selection process \cite{Antal2009, Tarnita2009}.
In this distribution, we compare the average abundance (i.e., frequency) of strategy $A$, $\langle x_A\rangle$, with that of $B$, $\langle x_B\rangle$.
If $\langle x_A\rangle > \langle x_B\rangle$, strategy $A$ prevails over (dominates) $B$.
Otherwise, $B$ prevails over $A$.
We derive the conditions for strategy dominance, i.e., the conditions which lead to $\langle x_A\rangle > \langle x_B\rangle$ or $\langle x_B\rangle > \langle x_A\rangle$.
To make progress, we consider weak selection (i.e., $0<\beta \ll 1$) \cite{Nowak2004, Traulsen2007, Wu2010, Tarnita2011, Allen2017}, under which individuals switch strategies with a nearly constant probability.
Weak selection may arise for the following reasons:
(i) individuals may be insensitive to the payoff-aspiration difference;
(ii) payoffs for different strategies are close; (iii) individuals are confused about the payoffs as they update strategies \cite{Tarnita2011}.

\section{Results}
Let us first consider personalized aspirations.
Given our assumptions, we calculate the average frequency difference between strategy $A$ and $B$, $\langle x_A-x_B\rangle$, in the stationary regime.
If $\langle x_A-x_B\rangle > 0$, strategy $A$ prevails over $B$; otherwise, strategy $B$ prevails over $A$.
Under weak selection, we find that strategy $A$ prevails over $B$ if $a+b>c+d$, and strategy $B$ prevails over $A$ if $a+b<c+d$.
This result holds for all weighted graphs without self-loops, for all distributions of aspirations, and for arbitrary number of updating functions.
Furthermore, if strategy $A$ and $B$ are both best replies to themselves (i.e., $a>c$ and $b<d$), our result reduces to the classical concept of risk-dominance.
It indicates that under the limit of weak selection, aspiration dynamics always select the risk-dominant strategy, which has a larger basin of attraction.

Our work extends previous studies \cite{Du2014, Du2015, Wu2018} in the following important directions:
(i) non-regular and weighted graphs can be considered;
(ii) individualized updating functions can be incorporated.
Our result also generalizes previous findings:
(i) the selection of risk-dominant strategy on regular graphs \cite{Du2015} is generalized to non-regular and weighted graphs, arbitrary number of updating functions, and personalized aspirations.
(ii) the coefficients $\sigma_0$ and $\sigma_1$ in the condition for strategy dominance in \cite{Wu2018} are shown to be equal and they can be rescaled to one by dividing a positive term, which leads to the condition of risk-dominance.

For an intuitive understanding of our result, we offer the following explanations.
Under weak selection, the expected payoffs of playing strategy $A$ and $B$ are evaluated at the neutral ($\beta=0$) stationary distribution of the aspiration dynamics (see SI for details).
In this distribution, individuals update strategies independently, which makes their strategies uncorrelated.
Individual $l$ thus on average interacts with neighbors using strategy $A$ as many times as those using $B$.
This means that the expected payoffs of $l$ are $\pi_{l,A}=(1/2)(a+b)$ and $\pi_{l,B}=(1/2)(c+d)$ when it plays strategy $A$ and $B$, respectively.
If $\pi_{l,A}>\pi_{l,B}$, individual $l$ is more satisfied when it uses strategy $A$ and the switching rate from $A$ to $B$ is less than that from $B$ to $A$.
Note that $\pi_{l,A}>\pi_{l,B}$ is equivalent to $a+b>c+d$.
Therefore, individual $l$ is more likely to be an $A$-player if $a+b>c+d$.
Since the above logic applies to any individual, the condition $a+b>c+d$ actually makes all the individuals feel more satisfied when they play strategy $A$.
As a consequence, the average frequency of $A$-players in the population is greater than that of $B$-players.
Similarly, $a+b<c+d$ results in more satisfaction when individuals play strategy $B$, which makes the average frequency of $A$-players less than that of $B$-players.

Comparing our result with that in well-mixed populations (equivalent to a complete graph in our model) \cite{Du2015}, we show that population structure does not alter the condition for strategy dominance.
In other words, the condition for strategy dominance under aspiration dynamics is robust to the underlying population structure.
The robustness property has practical advantages on strategy selection \cite{Tarnita2009}: (i) for a fixed game, the predictions are the same for a large class of population structures; (ii) to tell the fate of strategies, the population can be assume to be well-mixed.

\begin{figure}
  \centering
  \includegraphics[width=0.8\textwidth]{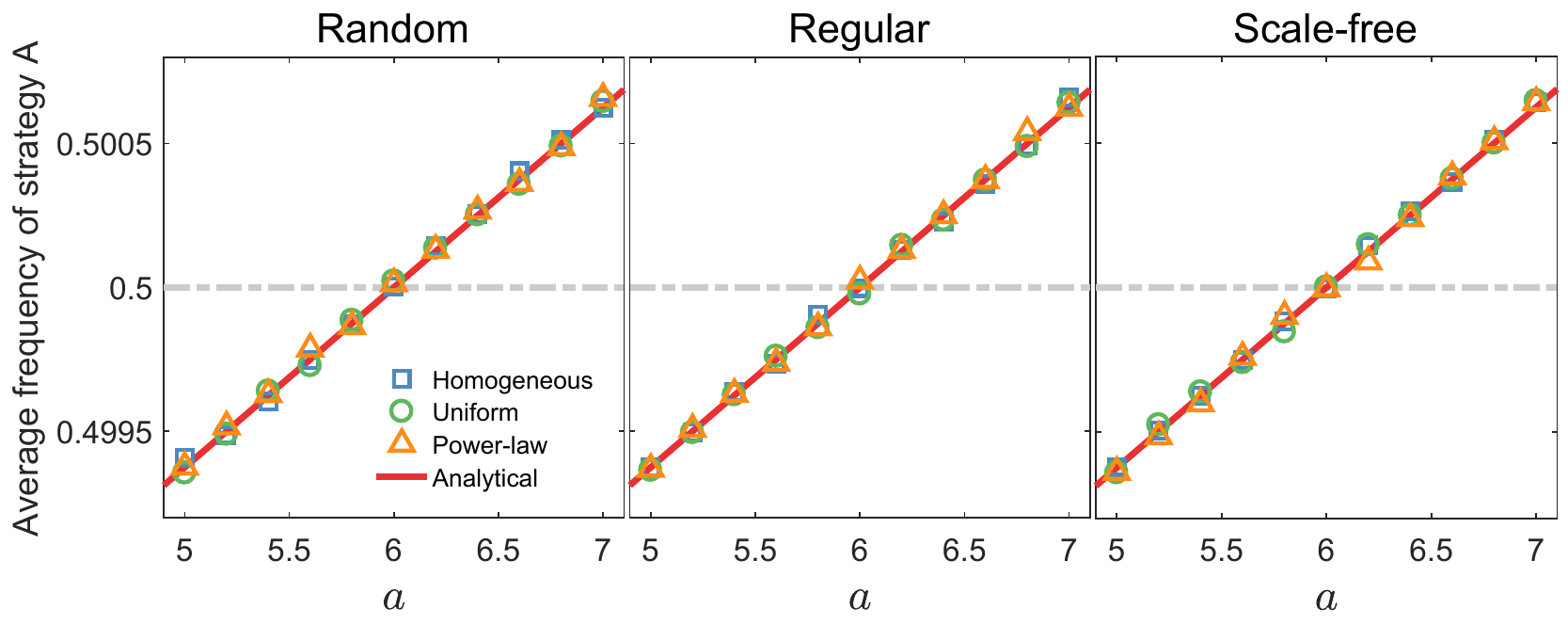}\\
  \caption{Aspiration dynamics generate robust predictions on weighted graphs.
  For the game, we set payoff value $b=0$, $c=5$, $d=1$ and leave $a$ as a tunable parameter.
  We plot the average frequency of strategy $A$, $\langle x_A \rangle$, as a function of $a$.
  Symbols represent simulation results while solid lines are analytical ones.
  We construct weighted graphs by first generating an undirected graph with average degree $\bar{k}$ and then assigning weights to edges.
  The undirected graphs considered are random graph (left), regular graph (middle), and scale-free network (right).
  For each type of network, we test three edge weight distributions:
  homogeneous---every edge has weight one (i.e., unweighted network);
  uniform---edge weights are uniformly selected from the integer set $\{1,2,3,4,5,6,7,8,9,10\}$;
  power-law---edge weights are randomly selected from the discrete power-law distribution (Zipf distribution with the exponent $s=3$).
  Each data point is obtained by averaging $\langle x_A \rangle$ in 200 independent runs.
  For each run, we calculate $\langle x_A \rangle$ by averaging the frequency of strategy $A$ in the last $1\times 10^7$ time steps after a transient time of $1\times 10^7$ time steps.
  Other parameters: $N=1000$, $\bar{k}=6$, $\alpha_l=2.0$ ($l=1,2,\cdots,N$), and $\beta=0.01$.
  }\label{fig2}
\end{figure}

Besides, by generalizing the Theorem in \cite{Wu2018} to non-regular graphs, our result is related to the structure coefficient $\sigma$ \cite{Tarnita2009, Nowak2010} derived for weak selection.
It is shown that $\sigma$ depends on the updating rule and the population structure (including the population size $N$).
But it does not depend on the payoff entries.
It summarizes the effect of population structure on the condition for strategy dominance.
For aspiration dynamics, we prove that $\sigma=1$ for a large class of population structures and there is no dependence on the population size.
This contrasts with the result obtained under imitation-based rules, which are shown to sensitively depend on the population structure and the population size \cite{Tarnita2009, Allen2017}.
In addition, $\sigma=1$ indicates that self-evaluation based rules does not lead to assortment of strategies for the purpose of strategy selection.
Therefore, evolutionary outcomes under self-evaluation based rules are intrinsically different from those under imitation-based rules.

When individuals' aspirations are contingent on their strategy in use (and thereby asymmetric), we find that the condition for strategy $A$ to be favored over $B$ under weak selection is $a+b> c+d - 2\left(\alpha_{B}-\alpha_{A}\right)$, where $\alpha_{X}$ is the aspiration of $X$-players ($X=A,B$).
Note that the condition now depends on aspirations.
Nonetheless, it is still robust to population structures, which generates invariant predictions for a large class of population structures.
Intuitively, the symmetry breaking of aspiration levels leads to additional asymmetry between strategy $A$ and $B$: $A$-players not only gain a different payoff but also have a different benchmark for satisfaction from $B$-players.
The expected level of satisfaction is now modified as $(1/2)(a+b)-\alpha_{A}$ and $(1/2)(c+d)-\alpha_{B}$ for $A$-players and $B$-players, respectively.
This modification alters the condition of risk-dominance derived under personalized aspirations and results in the dependence on aspiration levels.

So far, we only consider average payoffs.
Our framework also applies to accumulated payoffs (see SI for the conditions for strategy dominance).
We show that the condition of risk-dominance is invariant under accumulated payoffs, provided that aspirations are not contingent on strategies and the selection intensity is sufficiently weak.

\section{Conclusion and discussion}
In this work, we present a general framework to study aspiration dynamics in structured populations represented by graphs or networks.
We show that under weak selection, the condition for one strategy to be selected over the other is invariant on different population structures.
Moreover, this condition coincides with the condition of risk-dominance.
It indicates that aspiration dynamics always select the risk-dominant strategy.
When individuals' aspirations are contingent on strategies and thus asymmetric, the condition for strategy dominance is altered and determined by the difference between the aspirations of distinct strategies.
In this case, cooperators can prevail over defectors in the Prisoner's Dilemma if defectors aspire more than cooperators.

In addition to the irrelevance of social information, self-evaluation based updating rules have other features different from imitation-based rules: self-evaluation based rules are innovative \cite{Szabo2007}, which means they can revive strategies absent in the neighborhood without additional mechanisms such as random exploration or mutation; they increase their tendency to cooperate when more cooperators are present in the neighborhood (similar to conditional cooperators \cite{Fischbacher2001}) in Prisoner's Dilemma.
These features seem to be consistent with the recent findings on the possible features of human strategy updating \cite{Grujic2010, Gracia2012, Grujic2014, Rand2014}.
This suggests that self-evaluation based rules may be a good candidate for human strategy updating, which needs further empirical test.

Aspiration-based self-evaluation rules are also related to reinforcement learning.
The rationale behind reinforcement learning is the law of effect stated by Thorndike in 1898: actions bringing satisfactory effect will be more likely to be repeated and those leading to discomfort will be less likely to occur.
This is similar to our stochastic updating rules, except the reinforcement of actions \cite{Borgers1997, Erev1998}.
In practice, aspiration can evolve itself based on past experience \cite{Bendor2011}.
Although these features are not incorporated into our models, our work provides a first step towards evolutionary games in structured populations, whereas literature on reinforcement learning usually focuses on the simplest two-person repeated interactions (see a few recent exceptions on regular graphs \cite{Ezaki2016, Ezaki2017}).
Extending our model to incorporate aspiration adaptation and reinforcement of actions is a future direction.

For the effect of population structure on the fate of strategies, we show that aspiration dynamics generate quite different predictions from imitation-based ones.
When and how humans use one rule instead of another are yet to be explored.
A promising direction is to conduct experiments explicitly manipulating the information availability or monitoring the information request during the game \cite{Van2015, Mengel2018}.
Then, based on the distinct informational requirements of self-evaluation based and imitation-based rules, we may infer under what conditions human subjects tend to use these two classes of updating rules.
For theoretical studies, our work reveals a class of updating rules which generate invariant predictions for strategy dominance on a large class of population structures.
The reason may lie in the irrelevance of social peers' payoffs for strategy updating.
It remains unclear which other assumptions in updating rules affect the predictions on strategy dominance and what the most important ones are.
Future work along this line may lead to a deeper understanding about how updating rules alter the evolutionary outcomes, which may help to cultivate the optimal decision-making heuristics for the evolution of human cooperation.


\newpage
\setcounter{section}{0}
\renewcommand{\thefigure}{S\arabic{figure}}
\textbf{SUPPLEMENTARY INFORMATION (SI)}

\section{Model and notation}
\label{modelNotation}
\subsection{Population structure}
The population size is $N$ and the structure of the population is depicted by an undirected weighted graph with edge weights $w_{ij}$.
This graph is fixed during the whole evolution process.
Here, $w_{ij}>0$ denotes the weight of edge $ij$ while $w_{ij}=0$ means there is no connection between node $i$ and $j$.
Since edges are undirected, $w_{ij}=w_{ji}$.
Moreover, we define the weighted degree of node $i$ as $d_i=\sum_{j=1}^N w_{ij}$ \cite{Allen2017}.

\subsection{Discrete-time Markov chain with finite states}
In our model, individuals can play one of the two strategies, strategy $A$ and strategy $B$.
We consider the stochastic aspiration dynamics in structured populations.
The state of the system can be described by a (column) binary vector $\mathbf{s}=(s_1,s_2,\cdots,s_l,\cdots,s_N)^{\text{T}}$ where $\mathbf{s}(l)=s_l$ is the strategy of individual $l$: $s_l=1$ if individual $l$ uses strategy $A$, $s_l=0$ if it uses $B$.
We model the resulting evolutionary process as a discrete-time Markov chain $\{\mathbf{S}_n, n=0,1,2,3,\cdots\}$ with state space $\{\mathbf{s}_1, \mathbf{s}_2, \cdots, \mathbf{s}_M\}$ where $M=2^N$.
For each individual $l$, its strategy at the $k$-th step is thus $\mathbf{S}_k(l)$.
Note that throughout this SI text, all the vectors are column vectors by default.

\subsection{Payoff}
On the weighted graph, nodes are occupied by individuals and links indicate who interacts with whom.
Individuals interact with their nearest neighbors according to the game given by the following payoff matrix
\begin{align}
\bordermatrix{
  & A & B \cr
A & a & b \cr
B & c & d \cr},
\end{align}
and they collect the degree-weighted average payoff $\pi$.
Here, self-interactions are excluded and each individual must have at least one neighbor to interact with, which requires that $w_{ll}=0$ and $d_l>0$ for all $l$.

For each individual $l$, the number of times it interacts with an $A$-player at state $\mathbf{s}$ is $\sum_{k=1}^N w_{lk}s_k$ and that with a $B$-player is $\sum_{k=1}^N w_{lk}(1-s_k)$.
The payoff of individual $l$ at state $\mathbf{s}$ is thus
\begin{equation}\label{payoffCalculationAvg}
    \pi_l(\mathbf{s}) = \sum_{k=1}^N \frac{w_{lk}}{d_l}\left[a s_l s_k + b s_l (1-s_k) + c (1-s_l) s_k + d (1-s_l)(1-s_k)\right]
\end{equation}

\subsection{Aspiration-based updating functions}\label{AspUpdatingFunction}
In the real world, different individuals may make distinct decisions for the same shortfall of payoffs, which means they may have individualized ways of self-evaluation.
Such heterogeneity can be characterized by different updating functions \cite{Wu2018}.
We define individual $l$'s updating function as $g_l(u)$ ($l=1,2,\cdots,N$) and these functions represent the tendency to switch strategies.
For aspiration-based updating functions, $u=\beta(\alpha_l-\pi_l)$, where $\alpha_l$ is individual $l$'s aspiration level, $\pi_l$ its payoff, and $\beta>0$ is the intensity of selection \cite{Nowak2004, Allen2017}.
Weak selection means $\beta \ll 1$ and $\beta = 0$ is the neutral drift \cite{Tarnita2009}.
In addition, each function $g_l(u)$ should satisfy the following restrictions:
\begin{itemize}
  \item it is a probability, i.e., $g_l(u)\in[0,1]$ for $u\in (-\infty, +\infty)$;
  \item it is a strictly increasing function of $u$, i.e., $g'_l(u)=dg_l(u)/du>0$ for all $u$, which indicates that individuals with higher payoffs should have a lower tendency to switch;
  \item $g_l(0)>0$, which avoids frozen dynamics at the neutral drift.
\end{itemize}

\subsection{Transitions under aspiration dynamics}
Under aspiration dynamics, individuals' self-assessment of strategies triggers the transition between states.
At each time step, only one individual can update its strategy, either switch to the other strategy or keep the current strategy unchanged.
For the former, it means that the one-step transition between two distinct states of the associated Markov chain happens only if they possess the same entries everywhere except at one place which marks the strategy switching of the selected individual.
Formally, this means that the one-step transition between $\mathbf{s}_i$ and $\mathbf{s}_j$ ($i\neq j$) is possible only if $\sum_{l}|\mathbf{s}_i(l)-\mathbf{s}_j(l)|=1$.
Here, the operator $|\cdot|$ means to take the absolute value.
For convenience, we denote the set of states which can transit out of (or into) $\mathbf{s}_i$ as $\mathcal{N}_i$ ($\mathbf{s}_i \notin \mathcal{N}_i$), namely, the neighboring states of $\mathbf{s}_i$.
If the selected individual instead does not switch, the state of the Markov chain stays unchanged.
Therefore, the transition probabilities of the Markov chain are
\begin{equation}\label{TransitionAspDyn}
  p_{ij} =
      \begin{cases}
        & \frac{1}{N}g_{l_{ij}}(\beta(\alpha_{l_{ij}}-\pi_{l_{ij}}(\mathbf{s}_i))),~~~~~~~~~~~~~~~\text{ if }j\in \mathcal{N}_i, \\
        & 1 - \sum_{k\in \mathcal{N}_i}\frac{1}{N}g_{l_{ik}}(\beta(\alpha_{l_{ik}}-\pi_{l_{ik}}(\mathbf{s}_i))), \text{ if }i=j, \\
    & 0,  \text{otherwise},
    \end{cases}
\end{equation}
where $l_{ij}$ ($l_{ik}$) is the selected individual and it is uniquely determined by $\mathbf{s}_i$ and $\mathbf{s}_j$ ($\mathbf{s}_k$).
A simple way to locate $l_{ij}$ is by looking at the (unique) non-zero entry of vector $\mathbf{s}_j-\mathbf{s}_i$ when $\mathbf{s}_j \in \mathcal{N}_i$.
With all these transition probabilities, we have a transition matrix $\mathbf{P}$ whose $(i,j)$-th entry is $p_{ij}$, which describes the one-step transition probability between any states.

\section{Condition for strategy success}
\subsection{General condition}
Let us denote the number of individuals playing strategy $A$ at state $\mathbf{s}$ as $\phi(\textbf{s}) = \sum_{l=1}^N s_l$.
The abundance (frequency) of strategy $A$ and $B$ at state $\mathbf{s}$ is $x_A=\phi(\textbf{s})/N$ and $x_B=1-x_A$, respectively.
The frequencies of strategy $A$ at all states are denoted as a vector $\textbf{x} = (\frac{\phi(\mathbf{s}_1)}{N},\frac{\phi(\mathbf{s}_2)}{N},\cdots,
\frac{\phi(\mathbf{s}_{M-1})}{N},\frac{\phi(\mathbf{s}_M)}{N})^{\text{T}}$ and those of $B$ are $\textbf{1}-\textbf{x}$ where $\textbf{1}$ is a vector with all its elements being one.
Under aspiration dynamics, transition is possible between all the states (not necessarily in one step) and the number of steps taken can be odd or even.
This indicates that aspiration dynamics admit a unique stationary distribution $\textbf{u}$, in which each state occurs with a positive probability \cite{Grinstead2012} and all these probabilities sum up to one, i.e., $\textbf{u}^{\text{T}}\mathbf{1}=1$.
The stationary distribution $\textbf{u}$ can be obtained by solving a set of linear equations $\textbf{u}^{\text{T}} \textbf{P}=\textbf{u}^{\text{T}}$.
In this stationary distribution, we calculate the average abundance of strategy $A$ as $\langle x_A \rangle = \mathbf{u}^{\text{T}}\mathbf{x}$ and that of $B$ as $\langle x_B \rangle = \mathbf{u}^{\text{T}}(\mathbf{1}-\mathbf{x})=1-\langle x_A \rangle$, where the angle bracket $\langle \cdot \rangle$ means to take the average over all the states.

Strategy $A$ is more abundant than strategy $B$ if and only if $\langle x_A \rangle >\frac{1}{2}$ \cite{Tarnita2009, Tarnita2014}.
Since $\langle x_A \rangle + \langle x_B \rangle = \langle x_A + x_B\rangle = 1$, the condition $\langle x_A \rangle >\frac{1}{2}$ is equivalent to $\langle x_A\rangle > \langle x_B\rangle$ \cite{Nowak2010, Allen2014}, and further equivalent to $\langle x_A - x_B\rangle = \mathbf{u}^{\text{T}}(2\mathbf{x}-\mathbf{1})> 0$.
Note that $\textbf{u}$ and $\textbf{P}$ depend on the selection intensity $\beta$ and so do $\langle x_A \rangle$ and $\langle x_B \rangle$.
To show this dependence, we rewrite them as $\textbf{u}_{\beta}$, $\textbf{P}_{\beta}$, $\langle x_A \rangle_{\beta}$, and $\langle x_B \rangle_{\beta}$.
In the limit of weak selection $\beta \rightarrow 0$, we do a Talyor expansion of $\langle x_A - x_B \rangle_{\beta}$ with respect to $\beta$ at $\beta=0$ and get
\begin{equation}\label{avgAbundanceDiff}
  \langle x_A - x_B \rangle_{\beta}= \langle x_A - x_B \rangle_{0} + \langle x_A - x_B \rangle'_{0}\beta + O(\beta^2),
\end{equation}
where $\langle x_A - x_B \rangle'_{0}$ is the first-order derivative of $\langle x_A - x_B \rangle_{\beta}$ with respect to $\beta$, evaluated at $\beta=0$ and
$O(\beta^2)$ is the higher order terms of $\beta^2$, which is negligible as $\beta \rightarrow 0$.
Note that the first term in the right-hand side of (\ref{avgAbundanceDiff}) is the average abundance difference between strategy $A$ and $B$ at the neutral drift, which is zero \cite{Wu2018}.
Since $\beta > 0$, the condition for strategy $A$ to be more abundant than $B$ becomes
\begin{equation}\label{conditionDerivative}
  \langle x_A - x_B \rangle'_{0} = (\textbf{u}'_0)^{\text{T}}(2\textbf{x}-\textbf{1})>0.
\end{equation}

To get $\textbf{u}'_0$, we differentiate $\textbf{u}_{\beta}^{\text{T}} \textbf{P}_{\beta}=\textbf{u}_{\beta}^{\text{T}}$ with respect to $\beta$ and evaluate it at $\beta=0$, by rearranging items, we obtain
\begin{equation}\label{stationaryVectorDerivative}
   (\textbf{u}'_0)^{\text{T}}(\textbf{I}-\textbf{P}_0) = \textbf{u}_0^{\text{T}}\textbf{P}'_0,
\end{equation}
where $\textbf{I}$ is the identity matrix of dimension $M$, $\textbf{u}'_0=\frac{d}{d\beta}\textbf{u}_{\beta}|_{\beta=0}$, and $\textbf{P}'_0=\frac{d}{d\beta}\textbf{P}_{\beta}|_{\beta=0}$.
Note that $\textbf{I}-\textbf{P}_0$ is not invertible since it is not full rank.
Introduce the matrix $\textbf{W}=\textbf{1}\textbf{u}_0^{\text{T}}$, we know that the matrix $\textbf{I}-\textbf{P}_0 + \textbf{W}$ is invertible and its inverse, denoted as $\textbf{Z}_0$, is called the \textit{fundamental matrix} \cite{Grinstead2012}.
As shown in ref. \cite{Grinstead2012}, $\textbf{Z}_0=(\textbf{I}-\textbf{P}_0 + \textbf{W})^{-1}=\sum_{k=0}^{\infty}(\textbf{P}_0 - \textbf{W})^k=\textbf{I}+\sum_{k=1}^{\infty}(\textbf{P}_0^k - \textbf{W})$ and $(\textbf{I}-\textbf{P}_0)\textbf{Z}_0 = \textbf{I}-\textbf{W}$.
Note that $(\textbf{u}'_0)^{\text{T}}\textbf{W} = \textbf{0}^{\text{T}}$, right multiply $\textbf{Z}_0$ at both sides of equation (\ref{stationaryVectorDerivative}), we have
\begin{equation}\label{stationaryVectorDerivativeFinal}
   (\textbf{u}'_0)^{\text{T}} = \textbf{u}_0^{\text{T}}\textbf{P}'_0 \textbf{Z}_0.
\end{equation}

Substituting equation (\ref{stationaryVectorDerivativeFinal}) into (\ref{conditionDerivative}), we have that strategy $A$ is favored over strategy $B$ if
\begin{equation}\label{conditionFinalForm}
  \langle x_A - x_B \rangle'_{0} = \textbf{u}_0^{\text{T}}\textbf{P}'_0 \textbf{Z}_0(2\textbf{x}-\textbf{1}) =\textbf{u}_0^{\text{T}}\textbf{P}'_0 \left(\textbf{I}+\sum_{k=1}^{\infty}(\textbf{P}_0^k - \textbf{W})\right)(2\textbf{x}-\textbf{1}) = \textbf{u}_0^{\text{T}}\textbf{P}'_0 \textbf{c} > 0,
\end{equation}
where $\textbf{c}=\sum_{k=0}^{\infty}\textbf{P}_0^k(2\textbf{x}-\textbf{1})=(c_1,c_2,\cdots,c_i,\cdots,c_M)^{\text{T}}$.
In detail, $c_i = \sum_{k=0}^{\infty}\sum_{j=1}^M p_{ij}^{(k)}(\frac{2\phi(\mathbf{s}_j)}{N}-1)$ where $p_{ij}^{(k)}$ is the $(i,j)$-th entry of $\textbf{P}_0^k$.
Remind that $p_{ij}^{(k)}$ is the $k$-step transition probability from state $\mathbf{s}_i$ to $\mathbf{s}_j$.
Summing over all the possible states, $\sum_{j=1}^M p_{ij}^{(k)}(\frac{2\phi(\mathbf{s}_j)}{N}-1)$ is the average abundance difference between strategy $A$ and $B$ at the $k$-th step when starting at state $\mathbf{s}_i$.
Therefore, $c_i$ is the accumulated average abundance difference during the whole evolution when the initial state is $\mathbf{s}_i$.

Actually, equation (\ref{conditionFinalForm}) holds for any evolutionary dynamics which have a unique limiting stationary distribution and an equal abundance of strategy $A$ and $B$ at the neutral drift.
For example, the death-birth, birth-death process and pairwise comparison with symmetric mutations are of this class \cite{Ohtsuki2006, Traulsen2006b, Tarnita2009, Allen2014, Vasconcelos2017, Allen2017}.

\subsection{The condition for aspiration dynamics}

\subsubsection{Calculating the average accumulated abundance difference}\label{AvgAccAbundanceDiff}
Remind that $c_i = \sum_{k=0}^{\infty}\sum_{j=1}^M p_{ij}^{(k)}(\frac{2\phi(\mathbf{s}_j)}{N}-1)$ where $p_{ij}^{(k)}$ is the $k$-step transition probability from state $i$ to $j$ at the neutral drift (i.e., $\beta=0$).
It represents the accumulated average abundance difference during the whole evolution when starting at state $\mathbf{s}_i$.
Note that the summation over all the transition steps and that over all the states are interchangeable, we have $c_i = \sum_{j=1}^M \sum_{k=0}^{\infty} p_{ij}^{(k)}(\frac{2\phi(\mathbf{s}_j)}{N}-1)$ and and it is the average accumulated abundance difference between strategy $A$ and $B$ during the whole evolution.

Under aspiration dynamics, when $\beta=0$, the payoffs collected from the games do not affect individuals' strategy revisions.
More importantly, during the whole evolution, individuals' strategy revisions are completely independent of each other.
Due to this independence, we can study the much simpler two-state Markov chains for each individual instead of dealing with the whole chain with $2^N$ states.
For the two-state Markov chain of individual $l$, the state space is $\{0,1\}$ and the associated transition matrix is
\begin{align}
\bordermatrix{
& 1 & 0 \cr
1 & 1-\frac{1}{N}g_l(0) & \frac{1}{N}g_l(0) \cr
0 & \frac{1}{N}g_l(0) & 1-\frac{1}{N}g_l(0) \cr}.
\end{align}
The associated $k$-step transition probabilities are
\begin{eqnarray*}
  \mathrm{Prob}\left(\textbf{S}_k(l)=0|\textbf{S}_0(l)=0\right) &=& \mathrm{Prob}(\textbf{S}_k(l)=1|\textbf{S}_0(l)=1) = \frac{1}{2}\left[1+(1-\frac{2}{N}g_l(0))^k\right],\\
  \mathrm{Prob}(\textbf{S}_k(l)=1|\textbf{S}_0(l)=0) &=& \mathrm{Prob}(\textbf{S}_k(l)=0|\textbf{S}_0(l)=1) = \frac{1}{2}\left[1-(1-\frac{2}{N}g_l(0))^k\right].
\end{eqnarray*}

Since each individual's behavior is highly independent of each other, we can calculate their contribution to the average accumulated abundance difference separately.
Starting at state $\mathbf{s}_i$, at step $k$, the contribution of individual $l$ to the abundance difference is
\begin{eqnarray*}
  & &\left(\frac{2\mathbf{s}_i(l)}{N}-\frac{1}{N}\right)
\mathrm{Prob}(\textbf{S}_k(l)=\mathbf{s}_i(l)|\textbf{S}_0(l)=\mathbf{s}_i(l))
+ \left(\frac{1}{N}-\frac{2\mathbf{s}_i(l)}{N}\right)
\mathrm{Prob}(\textbf{S}_k(l)=1-\mathbf{s}_i(l)|\textbf{S}_0(l)=\mathbf{s}_i(l))\\
    & &=\frac{2\mathbf{s}_i(l)-1}{2N}
\left[1+(1-\frac{2}{N}g_l(0))^k\right]
+ \frac{1-2\mathbf{s}_i(l)}{2N}
\left[1-(1-\frac{2}{N}g_l(0))^k\right] = \frac{2\mathbf{s}_i(l)-1}{N}
\left(1-\frac{2}{N}g_l(0)\right)^k.
\end{eqnarray*}
By definition of the updating functions, $0<g_l(0)< 1$ for all $l$.
The contribution of individual $l$ to the accumulated abundance difference during the whole evolution is thus
\begin{equation*}
  \sum_{k=0}^{\infty}\frac{2\mathbf{s}_i(l)-1}{N}
\left(1-\frac{2}{N}g_l(0)\right)^k = \frac{2\mathbf{s}_i(l)-1}{N}\frac{N}{2g_l(0)} = \frac{2\mathbf{s}_i(l)-1}{2g_l(0)}.
\end{equation*}
Summing up all the individual contributions, we obtain the average accumulated abundance difference of the whole Markov chain as
\begin{equation}\label{coefficientCi}
  c_i = \sum_{l=1}^N \frac{2\mathbf{s}_i(l)-1}{2g_l(0)}
\end{equation}
By this formula, we obtain that $\mathbf{c}$ only depends on the updating functions.
It does not depend on the game, the aspiration levels, and the population structure.

\subsubsection{Deriving the exact formula}\label{ExactConditionAspDyn}
Denote $h(\textbf{s}_i)=\sum_{j=1}^M p'_{ij}c_j$, which is the $i$-th row of $\textbf{P}'_0 \textbf{c}$.
Equation (\ref{conditionFinalForm}) implies that the condition for strategy $A$ to be favored over $B$ is
\begin{equation}\label{conditionFinalFormSimplified}
  \langle x_A - x_B \rangle'_{0} = \langle h(\textbf{s}) \rangle_{0} > 0,
\end{equation}
where the bracket $\langle \cdot \rangle_{0}$ means to take the average over the neural stationary distribution (i.e., when $\beta=0$).

Substitute formula (\ref{coefficientCi}) of $c_j$ into $h(\textbf{s}_i)$, we have
\begin{equation}\label{coefficientH}
  h(\textbf{s}_i) = \sum_{j=1}^M p'_{ij}c_j = \sum_{j\neq i}^M p'_{ij}(c_j-c_i) = \sum_{j\neq i}^M p'_{ij}\sum_{m=1}^N \frac{\mathbf{s}_j(m)-\mathbf{s}_i(m)}{g_m(0)}.
\end{equation}
Note that the transitions can happen if $\textbf{s}_i$ and $\textbf{s}_j$ ($i\neq j$) differ at only one individual's strategy.
Combining with the transition probability (\ref{TransitionAspDyn}), we have that for $j \neq i$,
\begin{equation*}
  p'_{ij}\sum_{m=1}^N \frac{\mathbf{s}_j(m)-\mathbf{s}_i(m)}{g_m(0)} =
      \begin{cases}
        & \frac{1}{N}\frac{g'_{l_{ij}}(0)}{g_{l_{ij}}(0)}(\alpha_{l_{ij}}-\pi_{l_{ij}}(\mathbf{s}_i))\left(1-2\mathbf{s}_i(l_{ij})\right),~~~~~~~~~~~~~~~\text{ if }j\in \mathcal{N}_i, \\
        & 0,~~~~~~~~~~~~~~~~~~~~~~~~~~~~~~~~~~~~~~~~~~~~~~~~~~~~~~~~~~~~~\text{otherwise}.
    \end{cases}
\end{equation*}
where $g'_{l_{ij}}(u)=d g_{l_{ij}}(u)/du$ (see Section \ref{AspUpdatingFunction}).
The summation over all neighboring states can also be rewritten as the summation over all the individuals, which leads to
\begin{equation*}
  h(\textbf{s}_i) = \frac{1}{N}\sum_{l=1}^N \frac{g'_l(0)}{g_l(0)}(\alpha_{l}-\pi_{l}(\mathbf{s}_i))\left(1-2\mathbf{s}_i(l)\right).
\end{equation*}
Here, we drop the subscript $l_{ij}$ since the summation goes over all the individuals.

Substituting the payoff formula (\ref{payoffCalculationAvg}) into the above equation and using $s_l^2=s_l$, we obtain that for any state $\mathbf{s}$,
\begin{equation}\label{coefficientHFinal}
  h(\textbf{s}) = \frac{1}{N}\sum_{l=1}^N \frac{g'_l(0)}{g_l(0)}\left\{(1-2s_l)\alpha_{l}+\sum_{k=1}^N \frac{w_{lk}}{d_l}\left[a s_l s_k + b s_l (1-s_k)-c (1-s_l) s_k - d (1-s_l)(1-s_k) \right]\right\}.
\end{equation}

Evaluating $\langle h(\textbf{s}) \rangle_0$ needs to analyze the correlation of strategies in the neutral stationary distribution.
Under aspiration dynamics, when the selection intensity $\beta=0$ (neutral drift), individuals' strategy updating does not dependent on the aspiration level, the payoff, and the population structure.
The transition probabilities between any two states are thus the same in both directions, which indicates that the transition matrix is symmetric, i.e., $\textbf{P}_0=\textbf{P}_0^{\text{T}}$.
By the uniqueness of the stationary distribution $\mathbf{u}_0$ and the property of the transition (stochastic) matrix $\textbf{P}_0\textbf{1}=\textbf{1}$, we have $\textbf{1}^{\text{T}}\textbf{P}_0=\textbf{1}^{\text{T}}$, which leads to $\mathbf{u}_0=2^{-N}\textbf{1}$ through normalization.
This results in that at the neutral stationary distribution, each individual plays strategy $A$ with probability one-half, i.e., $\langle s_l \rangle_0 = 1/2$ for all $l$.
Moreover, since individuals' strategy updating is independent of each other, $\langle s_l s_k \rangle_0 = \langle s_l \rangle_0 \langle s_k \rangle_0=1/4$ when $l\neq k$.
Based on these, we have that the correlation of strategies at the neutral stationary distribution
\begin{equation}\label{correlationNeutral}
  \langle s_l s_k \rangle_0 = \frac{1+\delta_{lk}}{4},
\end{equation}
where $\delta_{lk}=0$ if $l\neq k$ and $\delta_{lk}=1$ if $l=k$.
Equation (\ref{correlationNeutral}) indicates that aspiration dynamics do not lead to assortment of strategies in the neutral stationary distribution.

Combining Equations (\ref{conditionFinalFormSimplified}), (\ref{correlationNeutral}), and (\ref{coefficientHFinal}), we obtain that
\begin{equation}\label{conditionFormulaFinalGeneral}
  \langle x_A - x_B \rangle'_{0} = \frac{1}{4N}\left(\sum_{l=1}^N \frac{g'_l(0)}{g_l(0)}\right)(a+b-c-d).
\end{equation}
Since $g_l(0)>0$, $g'_l(0)>0$ by definition, strategy $A$ outcompetes $B$ in abundance if $a+b>c+d$ and strategy $B$ outcompete $A$ if $a+b<c+d$.

Moreover, the average abundance of strategy $A$ in the stationary distribution can be obtained by combining equation (\ref{conditionFormulaFinalGeneral}) and the identity $\langle x_A + x_B \rangle_{\beta} = 1$, which leads to
\begin{equation*}
  \langle x_A \rangle_{\beta} = \frac{1}{2}+\frac{1}{8N}\left(\sum_{l=1}^N \frac{g'_l(0)}{g_l(0)}\right)(a+b-c-d)\beta + O(\beta^2).
\end{equation*}
\subsection{Asymmetric aspirations}
For aspirations contingent on the strategy in use, individuals playing strategy $A$ have  aspiration level $\alpha_{A}$ while those using $B$ have $\alpha_{B}$.
As shown in Section \ref{AvgAccAbundanceDiff}, coefficients $\mathbf{c}$ in Equation (\ref{conditionFinalForm}) is evaluated at $\beta=0$.
The asymmetry of aspirations thus does not affect $\mathbf{c}$.
However, $h(\mathbf{s})$ in Section \ref{ExactConditionAspDyn} now depends on both $\alpha_{A}$ and $\alpha_{B}$, and Equation (\ref{coefficientHFinal}) is modified as
\begin{equation}\label{coefficientHAsyAsp}
  h(\textbf{s}) = \frac{1}{N}\sum_{l=1}^N \frac{g'_l(0)}{g_l(0)}\left\{(1-s_l)\alpha_{B}-s_l\alpha_{A}+\sum_{k=1}^N \frac{w_{lk}}{d_l}\left[a s_l s_k + b s_l (1-s_k)-c (1-s_l) s_k - d (1-s_l)(1-s_k) \right]\right\}.
\end{equation}
Note that at the neutral stationary distribution, strategy correlations $\langle s_l s_k \rangle_0$ are independent of aspirations.
Combining Equations (\ref{conditionFinalFormSimplified}), (\ref{correlationNeutral}), and (\ref{coefficientHAsyAsp}), we obtain that
\begin{equation}\label{conditionFormulaFinalGeneralAsyAsp}
  \langle x_A - x_B \rangle'_{0} = \frac{1}{4N}\left(\sum_{l=1}^N \frac{g'_l(0)}{g_l(0)}\right)(a+b-c-d+2\alpha_B-2\alpha_A).
\end{equation}
Therefore, under the limit of weak selection, the condition for strategy $A$ to prevail over $B$ for asymmetric aspirations is
\begin{equation*}
  a + b > c + d - 2(\alpha_B-\alpha_A),
\end{equation*}
and that for $B$ to prevail $A$ is $a + b < c + d - 2(\alpha_B-\alpha_A)$.

In the Prisoner's Dilemma game (i.e., $a<c$ and $b<d$), $a+b$ is always less than $c+d$.
If the aspirations are symmetric, strategy $A$ is never favored since $a+b<c+d$.
With asymmetric aspirations, if $\alpha_B-\alpha_A$ is large enough, it is possible that $a + b > c + d - 2(\alpha_B-\alpha_A)$, which selects strategy $A$ over $B$.
Considering the evolution of cooperation, this means that cooperators can prevail over defectors if defectors aspire more than cooperators.
\subsection{Accumulated payoffs}
Under accumulated payoffs, the payoff of individual $l$ at state $\mathbf{s}$ is
\begin{equation}\label{payoffCalculationAcc}
    \pi_l(\mathbf{s}) = \sum_{k=1}^N w_{lk}\left[a s_l s_k + b s_l (1-s_k) + c (1-s_l) s_k + d (1-s_l)(1-s_k)\right].
\end{equation}
This leads to that
\begin{equation*}
  h(\textbf{s}) = \frac{1}{N}\sum_{l=1}^N \frac{g'_l(0)}{g_l(0)}\left\{(1-2s_l)\alpha_{l}+\sum_{k=1}^N w_{lk}\left[a s_l s_k + b s_l (1-s_k)-c (1-s_l) s_k - d (1-s_l)(1-s_k) \right]\right\}.
\end{equation*}
Since the way of payoff collection does not affect the correlation of strategies in the neutral stationary distribution, we obtain that
\begin{equation}\label{conditionFormulaFinalAcc}
  \langle x_A - x_B \rangle'_{0} = \frac{1}{4N}\left(\sum_{l=1}^N d_l \frac{g'_l(0)}{g_l(0)}\right)(a+b-c-d)
\end{equation}
and the condition for strategy $A$ to be favored over $B$ is $a+b>c+d$ and that for  strategy $B$ to be favored over $A$ is $a+b<c+d$.
This means that accumulated payoffs do not change the condition for strategy dominance under aspiration dynamics.

\end{document}